\documentclass[final,times]{elsarticle}

\usepackage{graphicx}

\usepackage{amsmath,amssymb}
\usepackage{amsthm}
\usepackage{booktabs}

\biboptions{square,round,sort&compress}

\journal{ArXiv}

\begin{document}

\begin{frontmatter}



\title{Modelling of immune cells as vectors of HIV spread inside a patient human body}


\author[label1]{Marcelo Margon Rossi\corref{cor1}}
\cortext[cor1]{Correponding author: Marcelo Margon Rossi. LIM01-HCFMUSP, University of S\~ao Paulo School of Medicine. Avenida Doutor Arnaldo, 455. 01246-903. S\~ao Paulo-SP, Brazil. FAX:+55 11 3061-7836.} \ead{mrossi@dim.fm.usp.br}
\author[label2,label3]{Luis Fernandez Lopez}
\address[label1]{LIM01-HCFMUSP, University of S\~ao Paulo, S\~ao Paulo, SP, 01246-903, Brazil}
\address[label2]{Dept. of Legal Medicine and Medical Ethics, University of S\~ao Paulo, S\~ao Paulo, SP, 01246-903, Brazil}
\address[label3]{CIARA, Florida International University, Miami, Fl, 33199, USA}

\begin{abstract}
The search to understand how the HIV virus spreads inside the human body and how the immune response works to control it has motivated studies related to Mathematical Immunology. Actually, researches include the idea of mathematical models representing the dynamics of healthy and infected cell populations and focusing on mechanisms used by HIV to invade target-host cells, viral dissemination (which leads to depletion of the T-cell pool and collapse of the immune system), and impairment of immune response. In this work, we show the importance of specific cells of immune response, as infection vectors involved in the dynamics of viral proliferation within an untreated patient, by using an ordinary differential equation model in which we considered that the virus infected target-cells such as macrophages, dendritic cells, and lymphocytes TCD4 and TCD8 populations. In conclusion, we demonstrate the importance of each cell-host and the threshold of viral establishment and posterior spread based on the presence of infected macrophages and dendritic cells in antigen-presenting processing, which leads to new infections (by wild or mutant virions) after immune response activation episodes. We presented an $R_0$ expression that provides the major parameters of HIV infection. Additionally, we suggest some possibilities of new targets for functional vaccines.
\end{abstract}


\end{frontmatter}


\section{Introduction}
\label{}
The epidemiology of infectious diseases has been studied for years and has been considered the control (and eradication) of infection dependent on the dynamics of epidemic process dynamics and human immunity \cite{tpk,dp,perelson}. The levels of immunization depend on vaccination use, the probability of occurrence of new cases and the metabolic consequences of an infected host. They are developed according to the level of complexity, the objective to be reached, the host-pathogen relationship and the population.\\
Based on the HIV infection process and dissemination, the range of maximum efficiency of a particular vaccine may not be reached \cite{walker,lzvl} because of the number of resistant strains that appear after treatment with antiretroviral drugs \cite{lmab,2cm} or by the formation of cellular reservoirs (macrophages and lymphocytes TCD4 cells chronically infected) \cite{hsbl,g3c}. Because viral secondary infections can occur in antigen presentation and lymphocyte activation or by macrophage activation in inflammation sites, a mathematical index expression considering these factors could help the effort to control and eradicate the disease more effectively.\\
A concept of great interest in epidemiology is the Basic Reproduction Number, $R_0$, which is the expected number of secondary cases per primary cases of infection in a completely susceptible population. This concept was initially proposed by MacDonald in the 1950s \cite{donalds}, when it was shown that for cases in which $R_0 >$1, the disease spreads into the population, and if $R_0 <$ 1, the disease will not progress. However, in a complex system, the deduction of simple expressions of $R_0$ is complicated due to the persistence of infection phenomena, such as various conditions and characteristics, the complexity level, the objective to be reached, and the nature of the host-pathogen relationship [3]. For these complex cases, the definition developed by Diekmann and Heesterbeek \cite{dh,dhm}, defined as Basic Reproduction Ratio, has been used to study the threshold criteria in heterogeneous systems. \\
In this study, we adapted the concept of $R_0$ to represent the phenomena of the HIV-1 infection and the dynamics of the immune system to control and eliminate it as an essential tool to test a hypothesis that results in more accurate outcomes. We applied the Next Generation Matrix Operator methodology to establish the threshold of viral infection based on immunological mechanisms, which is an important indicator of efforts to eliminate the infection.\\
\section{Description of the model}
\begin{figure}[H]
	\begin{center}
	\includegraphics[width=0.98\textwidth]{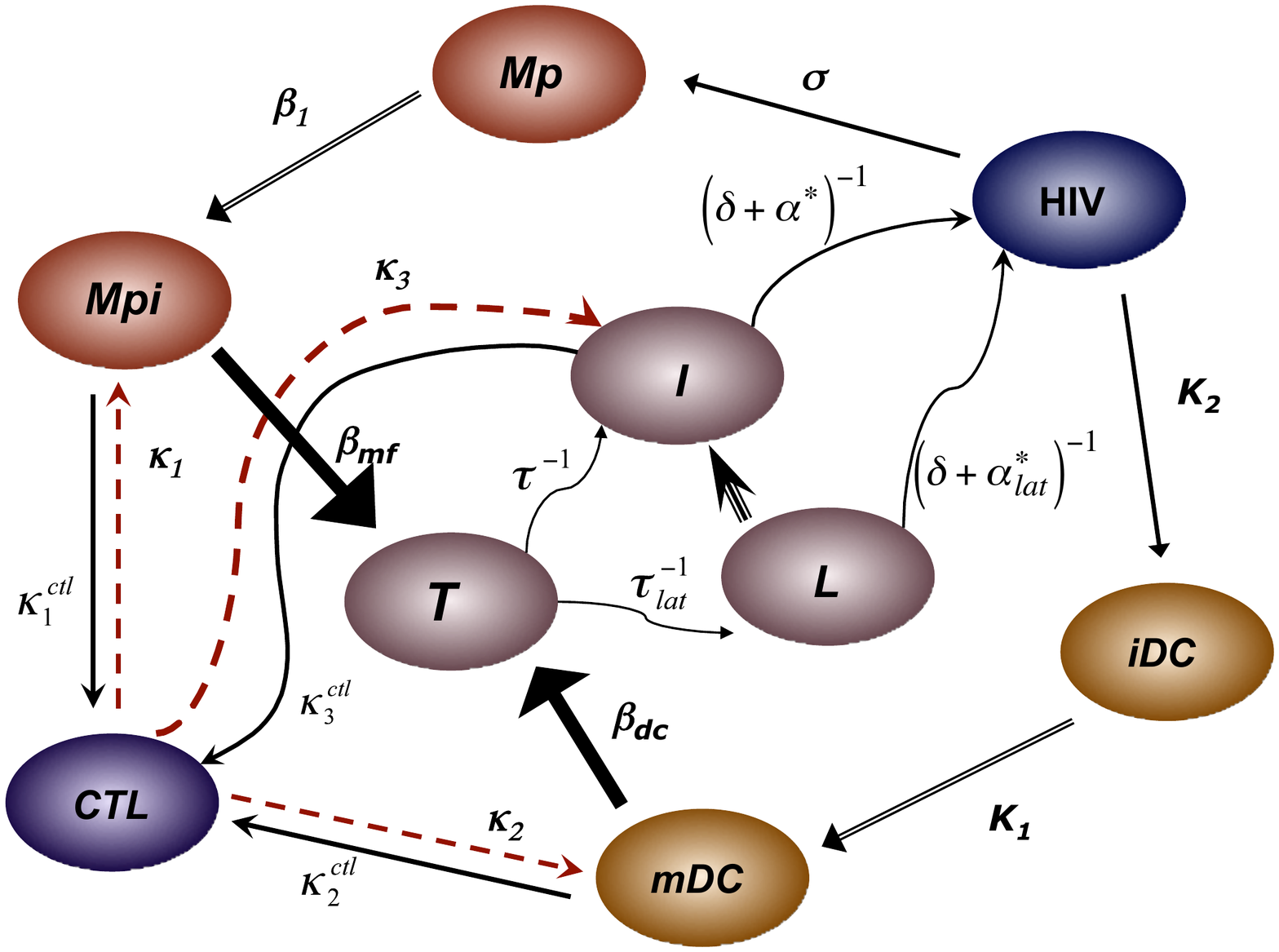}
	\end{center}
\caption{Immune system diagram showing the cell compartments involved in HIV-1 viral infection and the relationships among them used to represent parameters in the model \label{fig1}}
\end{figure}

The model description we propose is a realization of the immune system in which the various cell populations determine the set of equations \cite{wk2,kll}. The cell populations were $T$, the na夫e and not-infected TCD4+ lymphocyte population, $I$, the productively infected, and $L$, the latently (or chronically) TCD4+ infected lymphocyte subpopulations. The free viral load is represented by the $v$ equation. These equations consider additional cell populations that contribute to infection establishment. These cells are $Mp$, active not-infected macrophages; $Mpi$, active infected macrophages; $iDC$, immature not-infected dendritic cells and $mDC$ the infected and mature dendritic cells. The $CTL$ compartment of the model represents the cytotoxic action of TCD8+ lymphocytes. The effectors and the memory cytotoxic lymphocyte subpopulations were considered in the $CTL$ flux (Fig.~\ref{fig1}). All kinetic parameters representing the global flux are shown in Table 1.\\
The macrophage population is presented by (\ref{macrosus}), 
\begin{equation}
\frac{dMp}{dt}=s_{m}+l_{1}Mp-\delta_{f}Mp -\beta_{1}\frac{v^{n}}{v^{n}+K_{v}}Mp
\label{macrosus}
\end{equation}
Once the HIV virus penetrates the tissue barriers, macrophage sentinels find it and signal other immune cells. In (\ref{macrosus}), macrophages are recruited by a $s_m$ rate and activated according to the immune response level by $l_1$ rate. $\delta_f$ is the natural death rate. Once active macrophages contact viral particles, the particles are absorbed, and these absorption processes are represented by a saturation term (Holling II type), with a maximum rate parameter $\beta_1$ to lead to production of antigens. The exponent $n$ is a Hill coefficient and represents the possible cooperation among the HIV-1 molecular interactions with active macrophages. The viral particle ingestion initiates the active and infected macrophage population according to (\ref{macroinf}),
\begin{equation}
\frac{dMpi}{dt}=\beta_{1}\frac{v^{n}}{v^{n}+K_{v}}Mp-(\delta_{f}+\alpha')Mpi-\beta_{mf}Mpi\,T-k_{1}Mpi\,CTL \,.
\label{macroinf}
\end{equation}
These population dynamics show that infected macrophages do not suffer from the HIV-1 cytolytic effects as strongly as Tcells, and they can remain in the viral reservoir for a long period. According to Perno et al. \cite{pssp} and Aquaro et al. \cite{acbb}, virions may occur in macrophages with genetic sequences and different alternative splicing of those found in infected lymphocyte TCD4+ cells, showing a lesser genetic variance. These infected macrophages are given an additional death rate, $\alpha'$, that reflects possible HIV biochemical effects in cytosol, leading to an apoptosis state.  
The infected macrophages carry trapped viral particles to TCD4+ lymphocyte activation in the so-called "infectious synapse" process. These phenomena were quantified in the third term of (\ref{macroinf}), according to the Mass Action Law. $\beta_{mf}$ is the parameter encapsulating the contact and complex formation rate. The last term in (\ref{macroinf}) shows the cell-cell encounter rate displayed by the immune response. This immune action, related to the destruction of these infectious cells by cytotoxic Tcell lymphocytes (CTL), is proportional to the rate $k_1$. \\
During the HIV particle bypassing the tissue barriers, the recognizing process is activated, and the adaptive immune response initiates the presenting process by the dendritic cells. The $s_{id}$ parameter shows the cell influx into a specific tissue, and the increase of the active cell concentration (by adaptive immune response) is shown by the second term in (\ref{dendrimat}), with the $l_2$ parameter representing the activation rate.\\
\begin{equation}
\frac{diDC}{dt}=s_{id}+l_{2}\,iDC-\delta_{id}iDC-K_{1}\frac{v}{v+\kappa_v}\,iDC\,.
\label{dendrimat}
\end{equation}
in which dendritic cells die a natural death at the rate $\delta_{id}$. It was demonstrated that the relationship between the number of viral particles and mono-covalent connections to DC-membrane receptors follow a saturation process over time, according to Vanham et al. \cite{vpak} and Hlavacek et al. \cite{hsp,hwp}. The simpler mathematical expression used to represent these phenomena resembles the enzymatic rate by Michaelis-Menten. The process in which the virus has been holding onto the receptor is encapsulated in the third term of (\ref{dendrimat}), where $K_1$ is the maximum rate of the connected virus to cellular membrane receptors, $\kappa_v$ is the viral particle half-maximum concentration of dendritic cells. This formulation was chosen instead of Mass Action Law to exhibit the biochemical mechanisms described by Stebbing and Bower \cite{sb} and Wu and Kewalramari \cite{wk1}. \\
During this process, these cells mature (mDC) and process antigens into MHC class II complex, activating TCD4+ lymphocytes (\ref{denmature}). Therefore, the $mDC$ expression is \\
\begin{equation}
\frac{dmDC}{dt}=K_{1}\frac{v}{v+\kappa_v}iDC-\delta_{dc}\,mDC-\beta_{dc}mDC\,T-k_{2}\,mDC\,CTL \,.
\label{denmature}
\end{equation}
The number of contacts between these mature dendritic cells (with fragments or entire virus trapped) and na夫e TCD4+ lymphocytes is proportional to these two populations and has the probability $\beta_{dc}$ \cite{mdh}. These cells die at a rate $\delta_{dc}$ due to the death cell activation induced process \cite{h3p} and CTL lymphocyte cytotoxic action, at a rate $k_2$. The infected dendritic cell population directly migrates to the interior of the lymphoid organs to meet na夫e TCD4+ lymphocytes, activating and infecting it. The efficiency of infection in these lymphocytes is intimately associated with the viral integrity, an effective transfection process and adequate trapping in viral and Tcell binding \cite{cune,dpk,mptr}. \\
The TCD4+ lymphocyte population has a na夫e cell $s_4$ influx. These lymphocytes are activated after contact with HIV-infected macrophages or dendritic cells, which occurs during the antigen-presenting process and is increased by cloning expansion (\ref{cd4}). In active HIV-infected macrophages, this viral transport can occur in lymph organs or at inflammation sites \cite{psld}. The $\beta_{dc}$ and $\beta_{mf}$ parameters are contact rates that quantify the intensity of the adaptive immune response and the T-cell activation level by antigen processing \cite{vs}. Additionally, $\xi$ and $\phi$ parameters define, respectively, the infectious contact parameters of macrophages and dendritic cells in the antigen-presenting process. This antigen presentation was modeled by Queuing Theory, considering that the arriving process is close to a Poisson distribution's mean $1/c$ \cite{mcb}. The $c$ parameter shows the exponential decay life expectancy of free virus in blood. The survivor virus from immunological attack reaches the na夫e or active CD4 molecules with probability $[1-\exp(-ct)]/c$, with the parameter a representing the probability of the encounters to be infectious. Thus, the mathematical expression $\Psi(a,c;t)=a[1-\exp(-ct)]/c$ represents the efficiency of infectious contact between dendritic cells or macrophages, in the antigen presentation, and TCD4+ and CTL lymphocyte activation. The dynamics of na夫e TCD4+ lymphocytes follow the expression,\\
\begin{equation}
\frac{dT}{dt}=s_{4}+\left(\beta_{mf}Mpi+\beta_{dc}\,mDC \right)T-\Psi(a,c;t)\left(\xi Mpi+\phi \,mDC \right)T-\delta \,T\,.
\label{cd4}
\end{equation}
This mathematical formulation of an infectious encounter is very similar to the model used by McDonald in vector-host diseases \cite{donalds,mcb}. \\
After some na夫e lymphocytes are activated and infected, a proportion $f$ becomes productively infected TCD4+ lymphocytes and a proportion $(1-f)$ becomes chronically infected. It is assumed that this last T-cell group is activated at 0.0005 cell ($\mu l$.day)$^{-1}$ \cite{kp}, depending on the need to struggle with opportunistic infections, and it produces a virus in this process, as shown by (\ref{Tcellinf} and \ref{Tcellat}),\\
\begin{equation}
\begin{split}
\frac{dI}{dt}=f\,\Psi(a,c;t)\left(\xi Mpi(t-\tau)+\phi mDC(t-\tau) \right)T\left(t-\tau \right)e^{-(\delta + \alpha^*) \tau}\\
\quad  - (\delta + \alpha^{*})I+0.0005\,L - k_{3}I\,CTL \,,
\end{split} 
\label{Tcellinf}
\end{equation}
and,\\
\begin{equation}
\begin{split}
\frac{dL}{dt}=(1-f)\,\Psi(a,c;t)\left(\xi Mpi(t-\tau_{lat})+\phi mDC(t-\tau_{lat}) \right)T\left(t-\tau_{lat} \right)e^{-(\delta + \alpha_{lat}^{*}) \tau_{lat}}\\
\quad  -(\delta + \alpha_{lat}^{*})L - 0,0005L \,.
\end{split}
\label{Tcellat}
\end{equation}
Within these equations, $\tau$ and $\tau_{lat}$ parameters represent the period that HIV needs to complete its life cycle in a host cell. The infected T-cells die due to the HIV cytolytic action or the CTLﾕs immune response intensity. This occurrence is described by $\alpha^*$ and $\alpha^*_{lat}$ parameters and the cell life expectancy by ($\delta$+$\alpha^*$) and ($\delta$+$\alpha^*_{lat}$), which are the death rate parameters associated with productive and chronic cell populations, respectively. \\
The CTL cells play an important role in the immune process, signaling infected cells and destroying them by apoptosis mechanisms. Equation (\ref{ctl}) shows the dynamics of the CTL lymphocyte population on the immune system to control the spread of infection. The first term, $s_8$, represents the cell influx, and $\delta_{ctl}$ represents the natural death rate. $\xi'$ and $\phi'$ are parameters related to the activation of induced cell death of CTL by macrophages and dendritic cells, respectively. Each population cited previously contributes to immune response level through the rates $k_i^{ctl}$ ($i$=1,2,3), cooperating to increase the active effectors in the CTL population. \\
\begin{equation}
\begin{split}
\frac{dCTL}{dt}=s_{8}-\delta_{ctl}CTL- \left(\xi' Mpi+\phi' mDC \right) CTL\\
\quad +\left(k_{1}^{ctl}Mpi+k_{2}^{ctl}mDC+k_{3}^{ctl}I \right)CTL
\end{split} 
\label{ctl}
\end{equation}
Viral balance (\ref{vl}) shows the formation of free HIV-1 virus particles in the blood. The dynamics follow the expression,\\
\begin{equation}
\frac{dv}{dt}=Q_{1}(\delta + \alpha^{*})I+Q_{2}(\delta + \alpha_{lat}^{*})L+Q_{3}Mpi-K_{2}iDC\,v-\sigma\,\frac{v^{n}}{v^{n}+\kappa_v}Mp-c\,v \,.
\label{vl}
\end{equation}
where the two first terms represent the contributions from productively and chronically infected lymphocyte subpopulations, after cell death by HIV-1 cytolytic effects. The third term indicates the infected macrophage population effect on basal viral load. Parameters $Q_1$, $Q_2$ and $Q_3$ are the rates of participation of each infected cell population in HIV dissemination and viral pathogenesis. The other terms described as the free viral particles are withdrawn from the blood by dendritic cells (at rate $K_2$) and the numbers of HIV-1 virions that are encountered (and are captured) by active macrophages (at rate $\sigma$). $c$ is the viral clearance. These parameters evaluate the efficacy of the immunological actions.
All kinetic parameters were based on immunological and biochemical mechanisms of the immune response to the HIV infection process and were gathered from cited literature. The values are shown in Table~\ref{tabela1}. 
\begin{table}[h!]
	\centering
	\caption{\small Parameter values and biochemical meanings for parameter set from Equations~\ref{macrosus}-\ref{vl}}
	\small
	\label{tabela1}
	\begin{tabular}{p{0.40cm} c l c}
\toprule	
{\bf Parameter} & {\bf Value (unit)} &	{\bf Biochemical meaning} & {\bf References} \\ \midrule
$s_m$ &	16.3 cell($\mu$l.day)$^{-1}$ &	Active macrophage influx rate &	\cite{bh}\\
$l_1$ &	0.002 cell($\mu$l)$^{-1}$ &Activated macrophage growth rate & \cite{bh}\\
$\beta_1$ &	1.10	& Infected macrophage with viral particles  & \cite{dt,wk2}\\
$n$ & 0.55 & Hill coefficient & Adopted\\
$K_v$ & 	100 virions(ml)$^{-1}$&	Saturation coefficient &	Adopted\\
$\delta_f$ & 0.0285 (day)$^{-1}$ &	Activated macrophage natural death rate &	\cite{bh}\\
$\beta_{mf}$ &	4.50$\times$10$^{-04}$(day)$^{-1}$ &	Macrophage and CD4+ lymphocyte contact activation rate & \cite{vpak}\\
$\alpha'$ &	0.02 (day)$^{-1}$ &	Infected macrophage death rate due to HIV virus & \cite{dt}\\ 
$s_{id}$ & 	37.6 cell($\mu$l.day)$^{-1}$ &	Immature dendritic cell influx rate & \cite{mptr,kll}\\
$l_2$ &	0.085 cell($\mu$l)$^{-1}$	& Dendritic cell growth rate &	\cite{haase}\\
$K_1$ &	145&	Immature dendritic cell with viral particles & \cite{haase}\\
$\kappa_v$ &	4.5 & 	Complex Dendritic cell ﾐ HIV saturation constant &	Adopted\\
$\delta_{id}$ &	0.005 (day)$^{-1}$ &	Immature dendritic cell natural death rate	 & \cite{vpak}\\
$\beta_{dc}$	& 4.17$\times$10$^{-03}$(day)$^{-1}$ & Dendritic cell and CD4+ lymphocyte contact activation rate &	\cite{mpbg}\\
$\delta_{dc}$	& 0.025 (day)$^{-1}$ & 	Infected mature dendritic cell death rate &	\cite{dt}\\
$s_4$ &	0.025 cell($\mu$l.day)$^{-1}$ &Lymphocyte CD4+ influx rate &\cite{mptr,kll} \\
$a$ &	[0..1]	& Effective contact fraction among activated and infected cells \\	
$\xi$ &	0.0028(day)$^{-1}$ & 	Infected macrophage and TCD4+ contact rate &	Estimated \\
$\phi$ &	8.62$\times$10$^{-05}$(day)$^{-1}$&	Infected dendritic cell and TCD4+  contact rate &	Estimated\\
$\delta$	&0.015 (day)$^{-1}$&	Activated lymphocyte CD4+ natural death rate	& \cite{mpbg}\\
$f$ &	[0..1]&	Productively infected lymphocyte CD4+ fraction\\	
$\tau$ &	0.13 day	& Lymphocyte CD4+ intracellular viral production delay&	\cite{tpk}\\
$\alpha^*$ &	0.332 (day)$^{-1}$ &	Productively infected lymphocyte CD4+ additional death rate	& \cite{tpk,kll}\\
$\tau_{lat}$ &	20 day	& Viral production in latent lymphocyte CD4+ delay&	Adopted\\
$\alpha^*_{lat}$	& 0.132 (day)$^{-1}$ &	Latent lymphocyte CD4+ additional death rate	& \cite{mpbg}\\
$s_8$ &	0.0023 cell($\mu$l.day)$^{-1}$ &	CD8+ lymphocyte influx rate	& \cite{kll}\\
$\delta_{ctl}$ &	0.136 (day)$^{-1}$ &	CD8+ lymphocyte natural death rate &	\cite{mpbg}\\
$k_1$ &	1.50$\times$10$^{-03}$(day)$^{-1}$ & Infected macrophage and effector CTL contact rate &	\cite{aelm}\\
$k_2$ &	2.00$\times$10$^{-04}$(day)$^{-1}$ & Dendritic cell-HIV ﾒcomplexﾓ and effector CTL contact rate &	\cite{aelm}\\
$k_3$ &	2.50$\times$10$^{-04}$(day)$^{-1}$ & Infected CD4+ lymphocyte and effector CTL contact rate &	\cite{aelm}\\
$Q_1$ &	750 virions(cell)$^{-1}$ & Viral particle number per activated infected CD4+ lymphocyte &\cite{mpbg}\\
$Q_2$ &	80 virions(cell)$^{-1}$ &	Viral particle number per latently infected CD4+ lymphocyte &	\cite{mpbg}\\
$Q_3$ &	250 virions(cell)$^{-1}$ &	Viral particle number per infected macrophage	 & \cite{perelson}\\
$K_2$ &	0.15 ml(cell)$^{-1}$	& Absorption rate of HIV particles by immature dendritic cells & \cite{vpak}\\
$\sigma$ & 0.82 ml(cell)$^{-1}$ & Absorption rate of HIV particles by activated macrophages	& \cite{vpak}\\
$\xi'$ &	4.80$\times$10$^{-04}$ (day)$^{-1}$ &	Infected macrophage induced CTL death rate&	Estimated\\
$\phi'$ &	2.10$\times$10$^{-02}$ (day)$^{-1}$ & 	Mature dendritic cell induced CTL death rate  &	Estimated\\
$k_1^{ctl}$ &	1.50$\times$10$^{-03}$ cell($\mu$l.day)$^{-1}$ & CTL growth rate related to infected macrophage  & \cite{kll} \\
$k_2^{ctl}$ &	1.50$\times$10$^{-04}$ cell($\mu$l.day)$^{-1}$ & CTL growth rate related to mature dendritic cells  &\cite{kll} \\
$k_3^{ctl}$ &	6.1755$\times$10$^{-03}$ cell($\mu$l.day)$^{-1}$ & CTL growth rate related to infected CD4+ lymphocyte 	 & \cite{kll}\\
$c$ &	2.3 (day)$^{-1}$ &	Viral clearance rate	& \cite{perelson} \\ \bottomrule
	\end{tabular}
\end{table}
\subsection{Threshold calculation}
The necessity of considering other cellular reservoirs in addition to the TCD4+ lymphocytes (such as macrophages and dendritic cells) is very important for HIV spread containment and its complete control into the patientﾕs body, as well as the existence of a remarkable high efficiency of HIV vaccines. The model (\ref{macrosus}-\ref{vl}) represent, from a mathematical point of view, the immunological aspects of HIV infection, the growth and death rates, the number of contagious contacts (as the force of viral spread) and the "control" exerted by CTL populations on viral dissemination. This approach is very interesting to understand which factors in HIV infection mechanisms are important to initiate cell invasion, proliferation and dissemination, and to deplete TCD4+ populations, which cause impairment of the human immune system. \\
Equations (\ref{macrosus}-\ref{vl}) were conducted using Berkeley Madonna software 8.0.1, and simulation runs were performed with fourth order Runge-Kutta integration algorithm. The main objective was reaching the exact prediction of the HIV infection and disease evolution in an untreated patient. \\
\section{Results and Discussion}
The TCD4+ lymphocyte dynamics (Fig. 2) are similar to the classic HIV values found in the literature, and differences that appear could be due to the delays in the model. If we consider the biochemical events relative to the viral life cycle and the cell host half-life, this "time dilation" between the arrival and the ﾒcompletionﾓ of complete virions could be a type of immune system adaptation process to HIV-1. 
\begin{figure}[H]
	\begin{center}
	\includegraphics[width=0.98\textwidth]{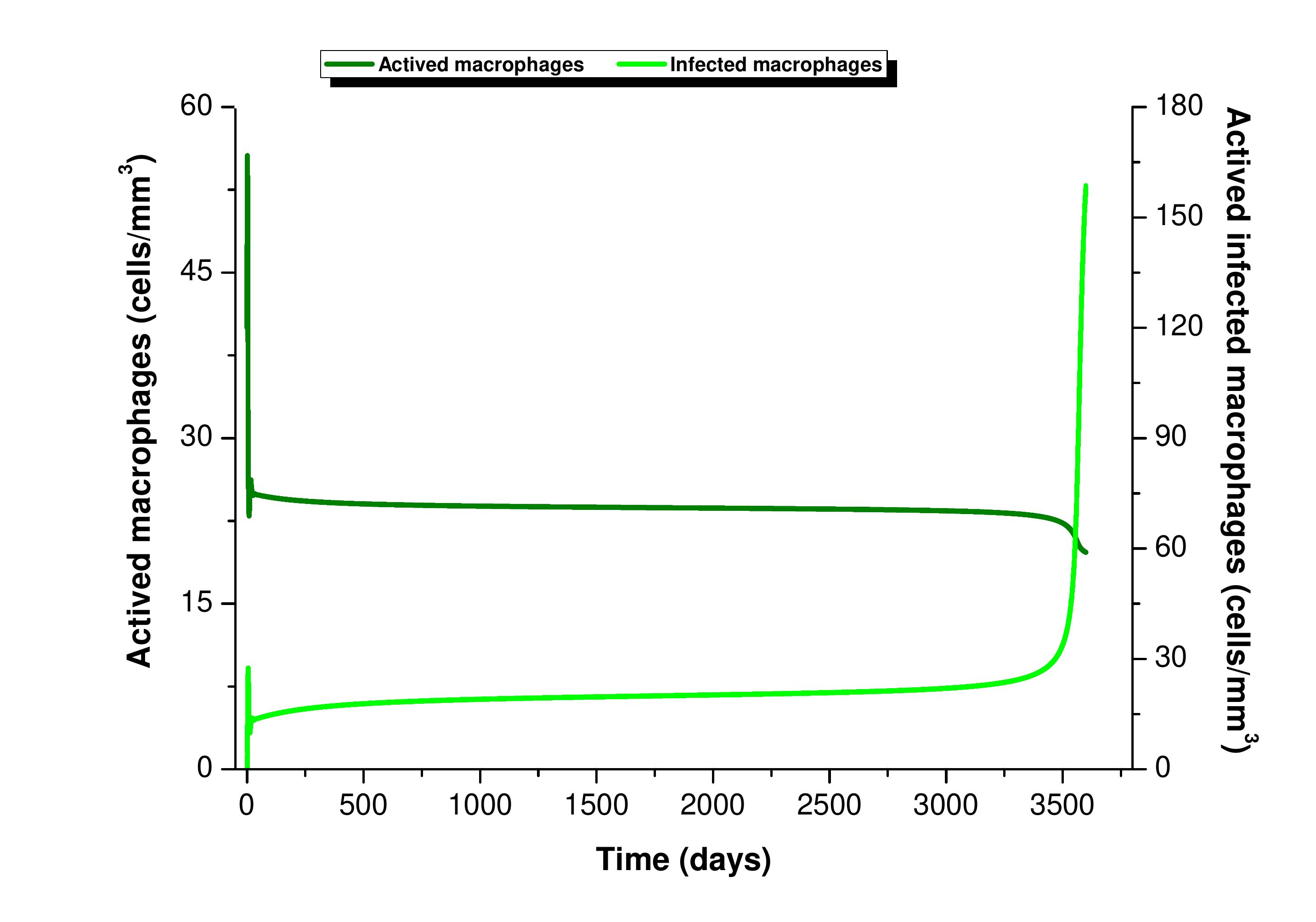}
	\end{center}
\caption{Simulation profiles of activated uninfected and HIV-infected macrophage populations, which show the immune transient response outcomes from viral contact in the early phase. The activated uninfected (dark green) and the HIV-infected macrophage cells (light green) show an early phase appears as an asymptotic phase and rises at the end of this period (which coincides with the AIDS period) \label{fig2}}
\end{figure}
In the early phase of the disease (0 to 60 days) (Fig. 2), approximately 6.0 \% of TCD4+ lymphocytes were infected. In the asymptomatic phase, this content was almost 8.4 \%. In chronically infected TCD4+ lymphocytes the content was 2.0 \% of the total lymphocyte population in the early phase and 0.15 \% in asymptomatic phase. However, if we consider the delayed input, the chronically infected TCD4+ lymphocytes play a specific (and unclear) role and act as a possible reservoir to HIV pathogenesis. The formation of the productively infected TCD4+ cells (Fig. 3) follows the profiles cited in specific HIV literature. In this figure, the value reaches 55 cells $\mu l^{-1}$ (in the early phase) and 65 cells $\mu l^{-1}$ during the asymptomatic phase and constant cytotoxic lymphocyte pressure. 
\begin{figure}[H]
	\begin{center}
	\includegraphics[width=0.98\textwidth]{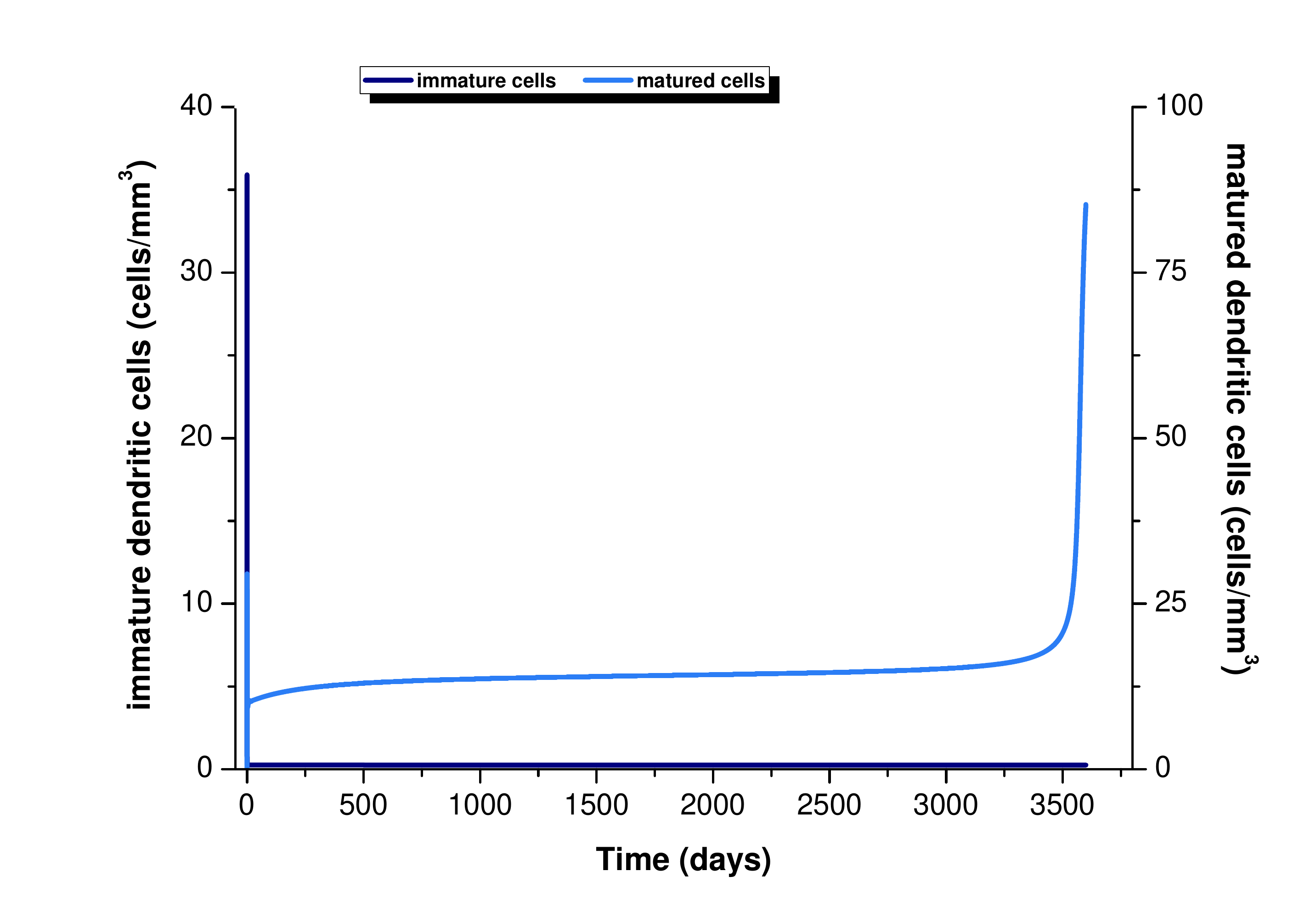}
	\end{center}
\caption{Simulation profiles of immature and mature dendritic cell populations, in which early phase appears as a rapid transient period as a result of HIV-1 contact and interaction with these cells in the antigen presentation process. The immature (dark blue) and the mature dendritic cells (light blue) show an initial and rapid transient phase, in which almost all dendritic cell are affected at the infection site, and it shows n plateau (on intermediate phase) resulting in an exponential growth of mature cells level trapping HIV in the AIDS phase. \label{fig3}}
\end{figure}
In evaluating the beginning of infection more carefully, the peak of infected cells reaches 100 cells $\mu l^{-1}$ in a time interval that is dependent or independent of delay. The infected lymphocytes show a death rate of approximately 0.0067 day$^{-1}$ due to CTL action. The relationship between the infected TCD4+ and the CTL content is approximately 18 \%, similar to [34].
\subsection{Macrophage cells}
\begin{figure}[H]
	\begin{center}
	\includegraphics[width=0.98\textwidth]{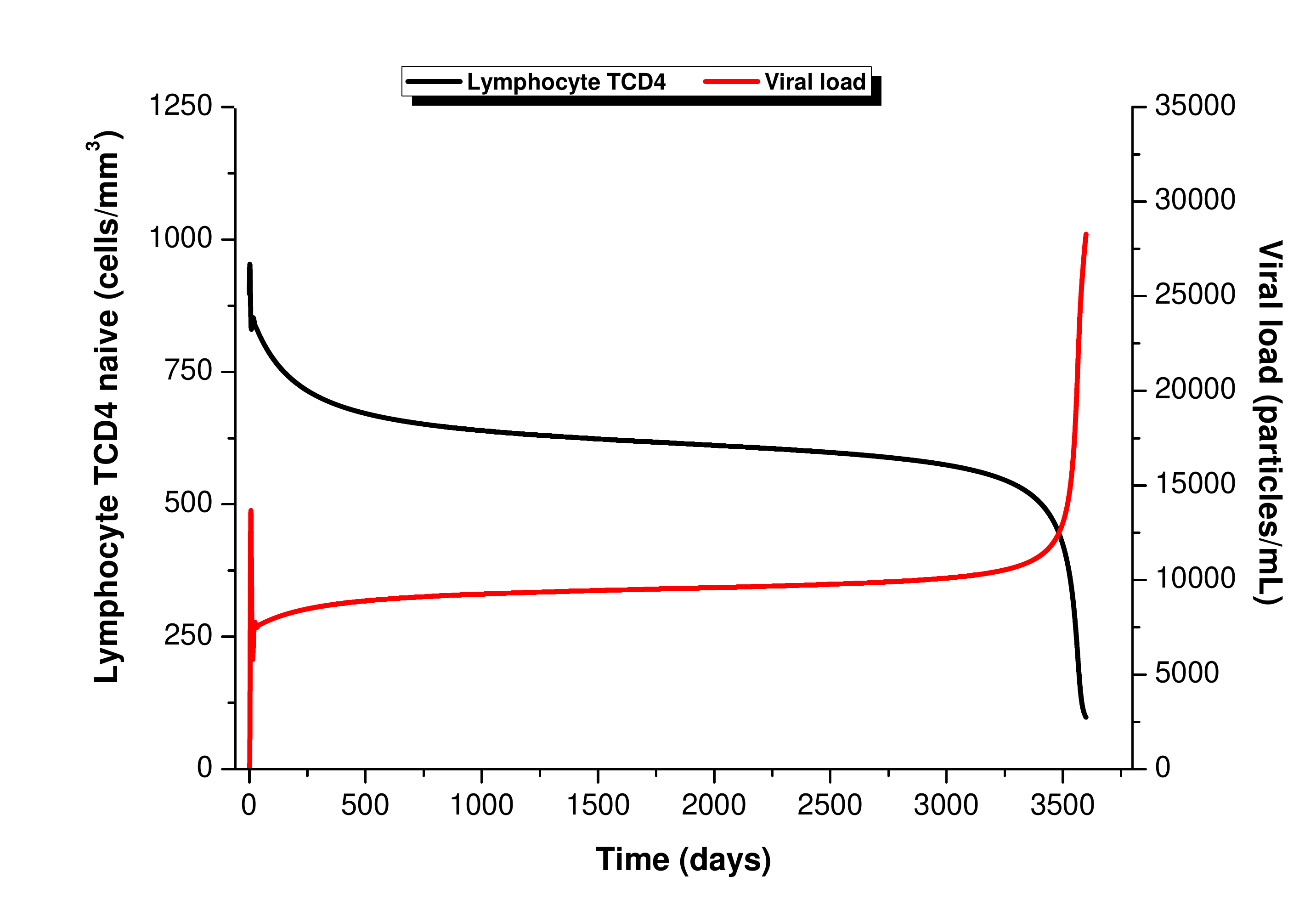}
	\end{center}
\caption{Simulation profiles of the viral load and productively HIV-infected TCD4+ lymphocyte population dynamics during the infection time. The TCD4+ lymphocyte (in black) population shows a transient period at an early phase during the initial immune response to HIV infection. Even as the viral load rises slowly and the infection proliferates, the Tcell populations deplete and the immune system is impaired. Viral load increases, spreading an exponential growth at the final phase, and rises to AIDS. \label{fig4}}
\end{figure}

When active macrophages meet viral particles, there is an increase in these cells at inflammation sites during early infection stages, and a small proportion suffer from infection. The mathematical simulations showed that 34 \% of total active macrophages become infected (almost two-fold the value found in [29] during the installation phase of HIV pathogenesis), and this level reaches 43 \% during the asymptomatic phase. The kinetics of macrophage proliferation show a pseudo-steady state in infected macrophage dynamics (Fig.~\ref{fig4}), reaching an average content of 18 cells $\mu l^{-1}$. Considering that nearly one-third of active macrophages become infected, its importance as a viral proliferation reservoir and infection maintenance is significant.\\
The dynamical equilibrium, as described in (\ref{ctl}), is modified by CTL lymphocyte population interference. The parameter $k_1$, representing the cytotoxic lymphocyte actions in the infected macrophage cell control, transfers a large sensibility to system equilibrium, but this control is not as pronounced in dendritic cells. All these characteristics reflect the control of infection dynamics.
\subsection{Dendritic cells}
The dendritic cells show a similar profile to the macrophage population. During the first 60 days (early infection phase), approximately 95 \% of mature dendritic cells have some viral particles trapped on its receptors. After an initial period, we can imagine that all dendritic cells derived from recognized HIV virions are infected. 
\begin{figure}[H]
	\begin{center}
	\includegraphics[width=0.98\textwidth]{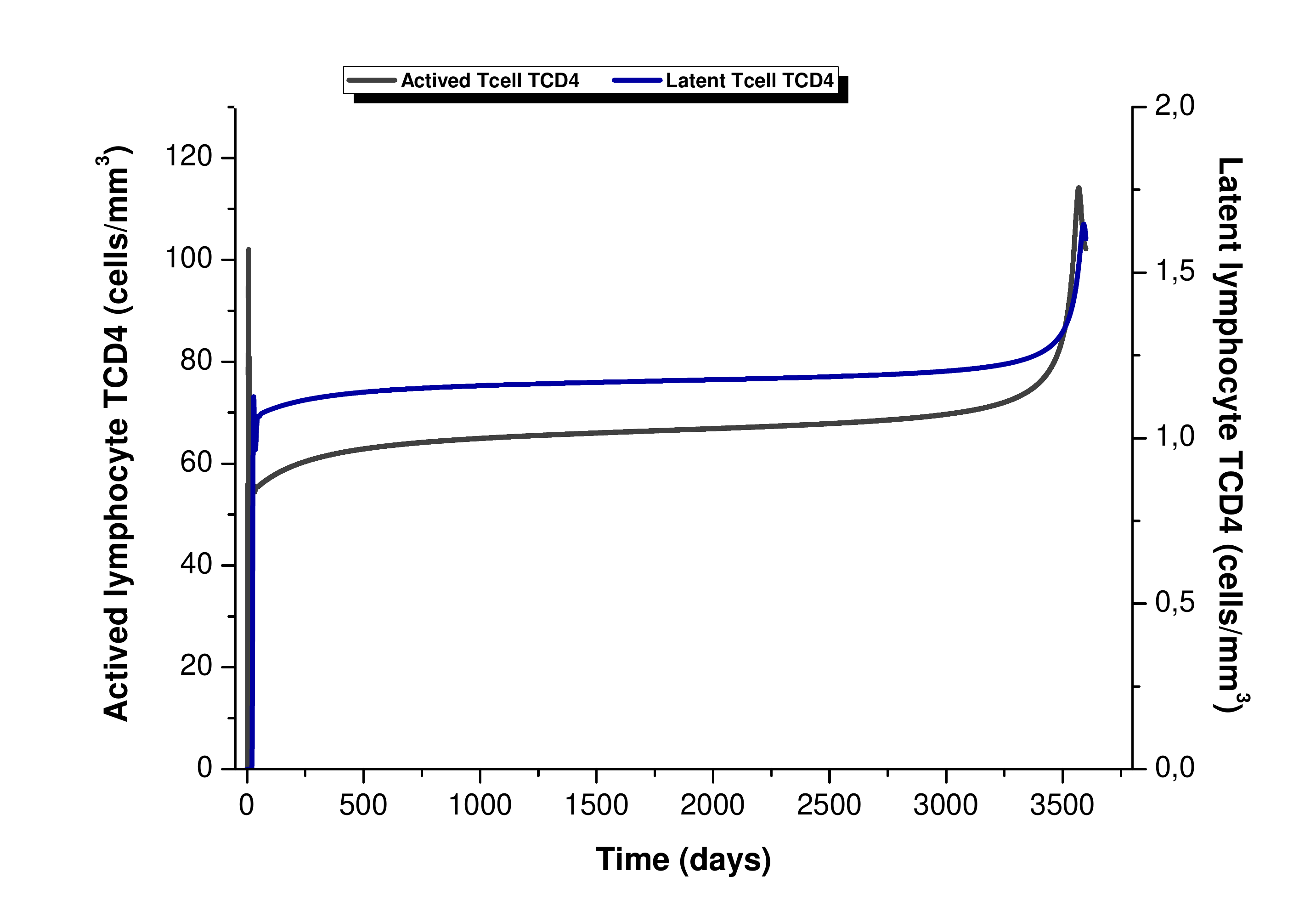}
	\end{center}
\caption{Simulation profiles of infected (in gray) and chronical infected (in blue) TCD4+ lymphocyte dynamics. The initial stage presents a transient period as a consequence of viral infectious process and immune responses, which goes on to a stable phase referred to as the asymptomatic phase, despite the fast growth due to intense viral proliferation and immune system depletion. The chronically infected Tcells (in blue) show similar dynamics with lower concentration. In the AIDS phase, these two cell sets grow fast resulting in immune system depletion. \label{fig5}}
\end{figure}
An equilibrium baseline is established (corresponding to the asymptomatic phase) at an average concentration of 15 cells $\mu l^{-1}$. In the last infection phase (Fig.~\ref{fig5}), there is an exponential increase of infected cells and viral proliferation rate [30]. In analyzing the kinetics parameter values and the simulation outcomes, the amount of virions absorbed per cell is 80 virions.cell$^{-1}$. This value is in agreement with the range of 100 to 180 particles per cell shown by Moris et al. [31] and Haase et al. [32]. The contact rate range between dendritic cells trapping virus and TCD4+ lymphocytes is 0.088 [15] and 0.0109 [31]. The value of saturation constant $\kappa_v$ was obtained from simulation outcomes.
\subsection{CTL lymphocytes}
The mathematical simulations show the three stages of viral infection (Fig.~\ref{fig4}): (a) the initial equilibrium perturbation as the viral cycle and infection is established; (b) the asymptomatic or chronic phase, in which a pseudo-steady state occurs in all cellular compartments and the virus spreads posteriorly, and (c) the final phase, in which the viral proliferation leads to depletion of the immune system. The CTL population is not sufficient to eradicate the HIV presence or the infection propagation because of either viral escape mechanisms or signaling cytokine depletion in the immunological response.\\
According to Rouzine et al. [33], there could be two possible mechanisms to control the HIV spread, and both relate to the viral load level. If the viral load is lower, the effort of CTL lymphocytes is TCD4+ active concentration dependent. A higher viral load is TCD4-independent but antigen-presenting dependent. This control mechanism can be verified in (\ref{ctl}) and results from a mathematical simulation, in which the TCD4/CTL ratio reaches values between 2.67 and 1.85. The viral escape phenomena, related to the inefficiency of CTL control, can be observed in Figs.~\ref{fig1} to \ref{fig5}.
\subsection{Infection level}
This study adapts the "force of infection" and "basic reproduction number" epidemiological methodologies, which are well established in epidemiological studies, as parameters of disease propagation in a susceptible population. In our study, the parameters represent the intensity of the antigen-presenting process, produced by infected macrophages and dendritic cells at infectious synapses, to na夫e TCD4+ lymphocytes in the adaptive immune response. After the pathogen-host complexes are formed, HIV is transferred to TCD4+ lymphocytes, where its genetic material is injected into host DNA, destroying it and infecting other susceptible cells. The progression of the disease is quantified by an index called Basic Reproduction Number, which corresponds to the average number of secondary cases that a unique infected case will cause in a susceptible population without treatment or any control of the infectious process [12]. 
\subsection{Basic reproductive number calculation}
After HIV penetrates the tissue barrier, it starts a cascade of cellular and biochemical mechanisms represented by the model (\ref{macrosus}-\ref{vl}). The efficiency of the humoral and adaptive response is represented in the matrix $\mathbf{\Sigma}$ (\ref{eq:matmorte}, \ref{NGOM}), which shows the probability of virus elimination (the first to fifth row) and the life expectancy of HIV in human blood (last vector in the form of the clearance rate $c$). The matrix $\bf \Lambda$ (\ref{eq:matent},~\ref{NGOM}) represents the death rates of infected cell populations, such as infected macrophages or mature dendritic cells and productively infected TCD4 lymphocytes. The last two vectors are zero because they are representatives of populations of CTL lymphocytes and viral load. Conceptually, the spectral radius $\rho$({\bf K}) is the largest absolute value of an eigenvalue of {\bf K} ({\bf K= -$\Sigma(s)\Lambda^{-1}$} where $\rho$({\bf K})) = sup {$|\lambda |$; $\lambda \in \sigma$({\bf K})) and is associated with $R_0$, according to (\ref{R0}). Mathematically, the spectral radius related to some positive matrix considers the expected numbers of offspring that an average individual would produce during its lifetime if the population lingers at low density. Each ($i,j$) entry of the Next Generation Matrix {\bf K} (\ref{eq:spectral}, \ref{NGOM}) is an expected number of secondary infections produced in compartment $i$ by a case initially in $j$. \\
This concept is related to the balance between the production of new infections by the group of infected or uninfected macrophages and dendritic cells on TCD4+ lymphocytes. It is also related to the control of viral proliferation (with their removal) generated by HIV-specific CTL response. Herein, the constant antigen presentations by HIV-infected macrophages impact new generations of viral infections and remain important reservoirs to the continuity of HIV proliferation and T-cell depletion. \\
By applying the next generation matrix methodology on model (\ref{macrosus}-\ref{vl}), the outcome is
\begin{equation}
R_0(\tau)= \frac{f\,\xi}{\delta+\alpha^*-0.0005}Mpi(\tau)+ \frac{(1-f)\,\phi}{\delta+\alpha^*_{lat}+0.0005}mDC(\tau).
\label{R0}
\end{equation}
Equation (\ref{R0}) shows that a possible target to the drug's action is focused on the $\tau$ term. The sensitivity is seen in~\ref{eq:index}, showing that the depletion profile of TCD4+ lymphocytes depends on the viral infection in macrophages and dendritic cells and the level of immunization. Additionally, in analyzing the destruction of all infected cells, such as HIV-infected macrophages, the variation of $R_0$ has a linear dependence on the death rate of these cells. Thus, although the target cells used by drug therapy are TCD4+ lymphocytes, the presence of dendritic cells and macrophages working like antigen-presenting cells still causes new infections, which maintain low levels of HIV virus. However, any inflammation process could trigger new HIV infection cases in lymphocytes resulting in viral load elevation and dissemination of the virus. \\
A sensitivity analysis in (\ref{R0}) of each parameter is shown by (and \ref{sensr0}),\\
\begin{subequations}
\label{eq:index}
\begin{equation}
\frac{\partial R_0}{\partial (\delta+\alpha^*)}=-f\,\frac{\xi\,Mpi(\tau)}{(\delta+\alpha^*)^2} \label{eq:sensprod}
\end{equation}
\begin{equation}
\frac{\partial R_0}{\partial (\delta+\alpha^*_{lat})}=-f\,\frac{\xi\,Mpi(\tau)}{(\delta+\alpha^*_{lat})^2} \label{eq:cron}
\end{equation}
\begin{equation}
\frac{\partial R_0}{\partial \xi}=-f\,\frac{\xi\,Mpi(\tau)}{(\delta+\alpha^*)-0.0005} \label{eq:macro}
\end{equation}
\begin{equation}
\frac{\partial R_0}{\partial \phi}=-f\,\frac{\xi\,mDC(\tau)}{(\delta+\alpha^*_{lat})+0.0005} \label{eq:dendrit}
\end{equation}
\begin{equation}
\frac{\partial R_0}{\partial \tau}=\frac{d\tau}{dt}\left[f\frac{\xi}{\delta+\alpha^*}\frac{\partial Mpi(\tau)}{\partial \tau}+(1-f)\frac{\phi}{\delta+\alpha^*_{lat}}\frac{\partial\,mDC(\tau)}{\partial\,\tau}\right] \label{eq:atraso}
\end{equation}
\end{subequations}
Equations~(\ref{eq:sensprod}) to (\ref{eq:atraso}) show the relationship between the $R_0$ index and the lymphocyte TCD4+ life expectancy, which notes that higher productively or chronically infected lymphocyte death rates are associated with lower rates of infection during the adaptive immune response. During the acute (or early) stage, cellular lyses induced by HIV occur and infected cells die by CTL action (\ref{eq:sensprod} and \ref{eq:cron}). Thus, if the cell-host dies, virus survivors are killed by the immune system. However, this dynamic equilibrium does not lower viral proliferation, and HIV-1 continues to infect and proliferate (under mutant particles form) escaping the immune system and infecting new target-cells. Thus, the vicious cycle restarts. \\
After the acute phase, the viral load is defined by the HIV replication rate and its clearance. Blips of intense viral replication are found during this period and are most likely caused by the inflammation process or opportunistic infections triggered by macrophages (or dendritic cells) and lymphocyte cell interaction. Hence, if HIV-infected macrophages (or dendritic cells) interact with na夫e TCD4+ and infect them, starting a new viral dissemination wave, the viral proliferation triggers other infected TCD4+ lymphocytes. This process is presented in (\ref{eq:macro}) and (\ref{eq:dendrit}). Specific studies \cite{g3c,acbb} cite that the viral population within the infected macrophages has a different mutant profile compared to that found in the infected lymphocyte subpopulation, probably due to different selective pressures to which HIV-1 is exposed. This information could explain the rapid dominance of wild-type virus after therapy cessation (Equations \ref{R0} and \ref{eq:index}).
\section{Conclusion}
Equation~\ref{eq:atraso} relates the impact of viral dissemination (e.g., the infection level) and the HIV replication cycle on infected macrophages and dendritic cells. It concludes that the interaction between the HIV virus and these host cells seems to have a profound influence on the exhaustion rate of the immune system.\\
The relations $f\, \Psi(a,c;t) \xi$ and $f\, \Psi(a,c;t) \phi$ show the efficiency of each contact between HIV-infected macrophages and dendritic cells, respectively, to lymphocyte T-cells in the antigen-presenting process. As the $a\xi$ and $a\phi$ terms represent the content of viral particles that infect the host cell and $f$ is the fraction of lymphocytes that are productively infected (the expression in parentheses represents the virus that survives the immune system attack), pre-exposure prophylaxis treatment therapies with fusion- or ligand-inhibitor drugs, for example, decrease the ﾒcontactﾓ terms $a\xi$ and $a\phi$. This process could be a barrier of viral entrance into the intracellular host environment. 
Infectious contacts caused by HIV-infected macrophages and dendritic cells lead to new infections in na夫e TCD4+ lymphocyte cells, influencing new immune responses to avoid viral proliferation, which could result in immune reconstitution inflammatory syndrome. The $R_0$ equation shows the mechanism of immune response that may relate to the targets of new drugs or new forms of pre-exposure prophylaxis treatment. The impact of new approaches to treatment can lead to the partial eradication of the disease and the formation of strains resistant to antiviral drugs by acting on cell tropism used by the viral subtypes present to infect. The model shows that populations of macrophages and dendritic cells constitute possible therapeutic targets to combat HIV/AIDS and targets for vaccination schemes.

\appendix

\section{The Next Generation Matrix Development}
\label{NGOM}
HIV-1 penetration through the tissue barrier starts a cascade of cellular and biochemical mechanisms attributable to this pathogen. The matrix ({\bf $M_2$ + D}) (\ref{eq:matmorte}) is composed of death rates of infected cell populations, such as infected macrophages, mature dendritic cells carrying viruses and productively infected lymphocytes TCD4 cells. The immune response depends on the humoral and adaptive response efficiency response, represented by matrix ({\bf $M_1$ + $\Sigma$}) (\ref{eq:matent}), which includes the probability of virus elimination (the first to fifth row) and the life expectancy of HIV in human blood (last vector in the form of the clearance rate $c$). The components of the last two vectors are zero because they are related to populations of CTL lymphocytes and viral load.\\
Mathematically, the spectral radius from a positive matrix is the expected number of offspring that an average individual would produce during its lifetime if the population decreases at low density. Each $i$ or $j$ entry of the matrix {\bf K} (\ref{eq:spectral}) is an expected number of secondary infections produced in compartment $i$ by a case initially in compartment $j$. Starting from the model described by equations {\bf A} to B, let $x(t)$ be the vector of populations and, assuming the TCD4+ lymphocyte population is large and the total number of cells infected at early stages of HIV infection is small, it is possible to assume that,
\begin{equation*}
x'(t)=T(s).x(t) - V.x(t)
\end{equation*}
The components of ({\bf $M_2$ + D}) and ({\bf $M_1$ + $\Sigma$}) matrices represent all rates that contribute to new infections at stage $j$ caused by contacts with an individual patient in stage $i$ and all death rates that contribute to decreasing the disease proliferation. {\bf D} is a positive diagonal matrix of removal (or death) rates, and {$\mathbf{\Sigma}$} represents all the removal rates caused by CTL lymphocyte control. When these two matrices are added, the matrix outcome is {\bf $M_2$ + D + $M_1$ + $\Sigma$} = {\bf V} (-{\bf V} is a Metzler matrix if {\bf V}= $m_{ij} \in \mathbb{R}_{nxn}$ all off-diagonal entries have non-negative values) (i.e., $m_{ij} >$ 0, $i \ne j$). \\
{\bf T}(s) is a non-negative transmission matrix, occurring by indirect form (by vector, intermediary host, contaminated food or environmental problem), horizontal direct form among individuals, vertical transmission (as mother-to-child infection type) and diagonal transmission. Thus, using an operator based in {\bf M}-matrix theory, another matrix {\bf K} can be associated with the -{\bf V} and {\bf T}(s) and,
\begin{equation*}
\mathbf{K}=-\mathbf{T}(s).\mathbf{V}^{-1}
\end{equation*}
Matrix {\bf K} can be interpreted as a "next-generation matrix" with respect to the infection process. The conditions of stability of this product matrix with Metzler matrices and threshold conditions of the system are calculated as the unique endemic equilibrium. In the last matrix, each element shows the particular contribution of each population to disease proliferation and cell death. In other words, each component is a type of "who infects who" point-of-view. Thus,\\
$\mathbf{M_2+D}$=
\begin{equation}
\begin{pmatrix}
{-(\delta_{f}+\alpha')}&{0}&{\ldots}&{}&{}\\ 
{0}&{-\delta_{dc}}&{0}&{\dots}&{}\\ 
{\dots}&{0}&{-(\delta+\alpha^{*})+0.0005}&{0}&{\dots}\\ 
{}&{\dots}&{0}&{-(\delta+\alpha^{*}_{lat})-0.0005}&{0}\\ 
{}&{}&{\ldots}&{0}&{0}\\ 
{}&{}&{}&{\ldots}&{0}
\end{pmatrix}
\label{eq:matmorte}
\end{equation}
and\\
$\mathbf{M_1+\Sigma}$=
\begin{equation}
\begin{pmatrix}
{0}&{\ldots}&{}&{}&{\ldots}&{\beta_{1} Mp}\\ 
{}&{0}&{\ldots}&{}&{\ldots}&{K_{1} iDC}\\ 
{}&{\ldots}&{0}&{\ldots}&{}&{}\\ 
{}&{}&{\ldots}&{0}&{\ldots}&{}\\ 
{k_{1}^{ctl}-\xi'}&{k_{2}^{ctl}-\phi'}&{k_{3}^{ctl}}&{0}&{-\delta_{ctl}}&{0}\\ 
{}&{}&{}&{\ldots}&{0}&{-c}
\end{pmatrix}
\label{eq:matent}
\end{equation}
The elements of {\bf T}(s) matrix are related to any entry that has any contribution to increase or elevate the cellular populations, as shown in (\ref{eq:matrixInfect}),\\
$\mathbf{T(s)}$=
\begin{equation}
\begin{pmatrix}
{-\beta_{mf}T_{0}}&{0}&{\ldots}&{}&{\ldots}&{0}\\ 
{0}&{-\beta_{dc}T_{0}}&{0}&{\ldots}&{\ldots}&{0}\\ 
{\Xi \,T(\tau)}&{\Phi\,T(\tau)}&{\Xi Mpi( \tau)}&{\Phi mDC( \tau)}&{0}&{\ldots}\\ 
{\Xi^* T(\tau_{lat})}&{\Phi^* T( \tau_{lat})}&{\Xi^*Mpi( \tau_{lat})}&{\Phi^* mDC( \tau_{lat})}&{0}&{\ldots}\\ 
{0}&{}&{\ldots}&{0}&{-(\xi' Mpi+\phi' mDC)}&{0}\\ 
{0}&{\ldots}&{}&{\ldots}&{0}&{-(\delta.Mp+K_{2}iDC)}
\end{pmatrix}
\label{eq:matrixInfect}
\end{equation}
where $\Xi= f\xi$, $\Xi^*= (1-f)\,\xi$, $\Phi= f\phi$ and $\Phi^*= (1-f)\,\phi$. \\
From the model (\ref{macrosus}-\ref{vl}) and ({\bf $M_2$ + D}) and ({\bf $M_1$ + $\Sigma$}) matrices, we could assume that this system has multiple discrete types of infected cell populations (macrophages, dendritic cells and lymphocytes). 
The Next Generation Matrix is defined as the square matrix {\bf K} (\ref{eq:spectral}) in which the $ij^{\,th}$ elements is the expected number of secondary infections of cell type $i$ caused by a single infected individual (e.g., infected cell or virus particle) $j$, assuming that the population from type $i$ is entirely susceptible. Thus, \\
$\mathbf{K}$=
\begin{equation}
\begin{pmatrix}
{-\frac{\beta_{mf}T_{0}}{\delta_{f}+\alpha^{'}}}&{0}&{\ldots}&{\ldots}&{0}&{\frac{\beta_{mf}T_{0}}{\delta_{f}+\alpha^{'}}\frac{\beta_{1}Mp_0}{c}}\\ 
{0}&{-\frac{\beta_{dc}T_{0}}{\delta_{dc}}}&{K_{23}}&{0}&{0}&{\frac{\beta_{dc}T_{0}}{\delta_{dc}}\frac{K_{1}iDC_0}{c}}\\ 
{\frac{\Xi T(\tau)}{\delta_{f}+\alpha'}}&{\frac{\Phi T(\tau)}{\delta_{dc}}}&{K_{33}}&{\frac{\Phi mDC(\tau)}{\delta +\alpha^*_{lat}+0.0005}}&{0}&{0}\\ 
{\frac{\Xi^* T(\tau_{lat})}{\delta_{f}+\alpha'}}&{\frac{\Phi^* T(\tau_{lat})}{\delta_{dc}}}&{K_{43}}&{\frac{\Phi^* mDC(\tau)}{(\delta + \alpha^{*}_{lat})+0,0005}}&{0}&{0}\\ 
{K_{51}}&{K_{52}}&{0}&{0}&{\frac{-(\xi' Mpi+\phi' mDC)}{\delta_{ctl}}}&{0}\\ 
{0}&{\ldots}&{}&{\ldots}&{0}&{\frac{-(\delta.Mp+K_{2}iDC)}{c}}
\end{pmatrix}
\label{eq:spectral}
\end{equation}
\\
where the $K_{23}, K_{33}$, $K_{43}$, $K_{51}$ and $K_{52}$ terms are\\
\begin{eqnarray}
K_{23}&:=&\frac{\beta_{dc}T_0}{(\delta + \alpha^* - 0.0005)}\left(\frac{\beta_1 (k_1^{ctl}-\xi')}{(\delta_f + \alpha')c}Mp_0 + \frac{K_1 (k_2^{ctl}-\phi')}{\delta_{dc}c}iDC_0 \right) \nonumber \\
K_{33}&:=&\frac{\Phi T(\tau)}{\delta+\alpha^*-0.0005}\left(\frac{\beta_1(k_1^{ctl}-\xi')}{(\delta_f + \alpha')c}Mp_0 + \frac{K_1(k_2^{ctl}-\phi')}{\delta_{dc}c}iDC_0\right)+\frac{\Xi Mpi(\tau)}{\delta+\alpha^*-0.0005} \nonumber \\
K_{43}&:=& \frac{\Phi^* T(\tau_{lat})}{\delta+\alpha^*-0.0005}\left(\frac{\beta_1(k_1^{ctl}-\xi')}{(\delta_f + \alpha')c}Mp_0+ \frac{K_1(k_2^{ctl}-\phi')}{\delta_{dc}c}iDC_0 \right)+\frac{\Xi^* Mpi(\tau_{lat})}{\delta+\alpha^*-0.0005} \nonumber \\
K_{51}&:=&\frac{\xi'\,CTL_{0}}{\delta_{f}+\alpha'}-\frac{\xi'Mpi+\phi'mDC}{\delta_{ctl}(\delta_f+\alpha')} \nonumber \\
K_{52}&:=& \frac{\phi'\,CTL_{0}}{\delta_{dc}}-\frac{\xi'Mpi+\phi'mDC}{\delta_{ctl}\delta_{dc}}(k^{ctl}_2-\phi').
\end{eqnarray}
\section{$R_0$ equation sensitivity analysis}
\label{sensr0}
The sensitivity to each of the parameters described in model (\ref{macrosus}-\ref{vl}) is obtained by the normalized forward sensitivity index. The generalized $S_p$ expression to a specific parameter $p$ is,
\begin{equation*}
S_p= \sum \frac{\partial R_0}{\partial p}\frac{p}{R_0}\frac{\partial p}{\partial t}\frac{t}{p}. 
\label{normalindex}
\end{equation*}
This mathematical relation was used to analyze cell contact, activation and death parameters. The $R_0$ relationship with productive and chronical infected TCD4+ lymphocyte death rate, the antigen activation process related to encounters between macrophages (or dendritic cells), HIV-infected and uninfected lymphocytes TCD4+, and its specific delay $\tau$ are shown in (\ref{infdeath}) and (\ref{latdeath}),
\begin{equation}
S_{1/(\delta + \alpha^*)}= \frac{t}{1+ \Theta(\tau)},
\label{infdeath}
\end{equation}
\begin{equation}
S_{1/(\delta + \alpha^*_{lat})} = \frac{t}{1+ \Theta^{-1}(\tau)} ,
\label{latdeath}
\end{equation}
\begin{equation}
S_{\xi}= \frac{\xi}{1+ \Theta(\tau)},
\label{xi}
\end{equation}
and
\begin{equation}
S_{\phi}=  \frac{\phi}{1+ \Theta^{-1}(\tau)}\, .
\label{phi}
\end{equation}
where
\begin{equation}
\Theta(Mpi, mDC; \tau)\equiv \Theta(\tau):=\frac{\phi}{\xi}\,\frac{mDC(\tau)}{Mpi(\tau)} \frac{(1-f)}{f} \frac{(\delta + \alpha^*)-0.0005}{(\delta + \alpha^*_{lat})+0.0005} \nonumber
\end{equation}
It is noted that, if infected lymphocyte population death rate increases by 1.0 \% due viral proliferation, the $R_0$ value increases nearly by 2.3 \% (\ref{infdeath}). If we consider that the index is related to chronically infected lymphocyte death rate, the $R_0$ variation is approximately lower, 0.015 \% (\ref{latdeath}).\\
In the absence of antiviral treatments, the variation of 1.0 \% in the parameter related to encounter rate ($\phi$ and $\xi$), results in a variation of nearly 0.3 \% (relative to macrophages) and 5.8\,$\times$\,10$^{-5}$ \% (relative to dendritic cells), respectively. It could indicate that macrophages population stresses higher $R_0$ than dendritic cells, conforming showed at (\ref{xi}) and (\ref{phi}).\\
The $R_0$ sensitivity is directly proportional to the complete viral life cycle over time and the macrophage and dendritic cell infected populations, as can be seen in (\ref{R0tau}),
\begin{equation}
S_{\tau}=\left[\frac{\partial }{\partial \tau}R_0 \right]\frac{\partial \tau}{\partial t}\frac{t}{R_0}.
\label{R0tau}
\end{equation}
{\bf{References}}



\begin{thebibliography}{00}


\bibitem{tpk}Thakar J, Pilione M, Kirimanjeswara G, Harvill E T, Albert R (2007)  Modeling system-level regulation of host immune response.  PLoS Comput Biol 3(6):e109.doi:10.1371/journal.pcbi.0030109.
\bibitem{dp}Dixit M N, Perelson A S (2004) Complex patterns of viral load delay under antiretroviral therapy: Influence of pharmacokinetics and intracellular delay. J Theor. Biol. 226:95-109.
\bibitem{perelson}Perelson A S (2002) Modeling viral and immune systems dynamics. Nature Rev. Immunol. 2:28-36.
\bibitem{walker}Walker B D (2006) Immune control and immune failure in HIV infection. Clin. Care Options.
\bibitem{lzvl}Lin M L, Zhang Y, Villadangos J A, Lew A M (2008) The cell biology of cross-presentation and the role of dendritic cell subsets. Immunol Cell Biol, doi:10.1038/icb.2008.3. 
\bibitem{lmab}Landi A, Mazzoldi A, Andreoni C, Bianchi M, Cavallini A et al. (2008) Modeling and control of HIV dynamics. Comp Meth Prog Biomed 89:162-168.
\bibitem{2cm}Cassels S, Clark S J, Morris M (2008) Mathematical models for HIV transmission dynamics. J Acquir Immune Syndr 47:S34-S39.
\bibitem{hsbl}Hladik F, Sakchalathorn P, Ballweber L, Lentz G, Fialkow M et al. (2007) Initial events in establishing vaginal entry and infection by human immunodeficiency virus type-1. Immunity 26, DOI: 10.1016/j.immuni.2007.01.007.
\bibitem{g3c}Gorry P R, Churchill M, Crowe S M, Cunningham A L, Gabuzda D (2005) Pathogenesis of macrophage tropic HIV-1. Curr. HIV Res. 3:53-60.
\bibitem{donalds}MacDonalds G (1952) The analysis of equilibrium in Malaria.  Trop Dis Bull 49:813-828.
\bibitem{dh}Diekmann O, Heesterbeek J A P (2000)  Mathematical epidemiology of infectious diseases: Model buildings, analysis, and interpretation. Chichester, NewYork. John Wiley Ed.
\bibitem{dhm}Diekmann O, Heesterbeek J A P, Metz J A J (1990) On the definition and the computation of the basic reproduction ratio R0 in models for infectious diseases in heterogeneous populations. J Math Biol 28:365-382.
\bibitem{pssp}Perno C F, Svicher, V, Schols D, Pollicita M, Balzarini J et al. (2006)  Therapeutics strategies towards HIV-1 infections in macrophages.  Antiviral Res 71:293-300.
\bibitem{acbb}Aquaro S, Calio R, Balzarini J, Bellocchi M C, Garaci E et al. (2002) Macrophage and HIV infection: Therapeutical approach toward this strategic virus reservoir. Antiviral Res 55:209-225.
\bibitem{vpak}Vanham G, Penne L, Allemeersch H, Kerstens L, Willems E et al. (2000)  Modeling HIV transfer between dendritic cells and Tcells: Importance of HIV phenotype, dendritic cell Tcell contact and Tcell activation.  AIDS 14:2299-2311.
\bibitem{hsp}Hlavacek W S, StilianakisN I, Perelson A S (2000)  Influence of follicular dendritic cells on HIV dynamics. Phil Trans R Soc Lon B Biol Sci 355:1051-1058.
\bibitem{hwp}Hlavacek W S, Wofsy C, Perelson A S (1999) Dissociation of HIV-1 from follicular dendritic cells during HAART: Mathematical analysis.  Proc Natl Acad Sci USA 96:14681-14686.
\bibitem{sb}Stebbing J, Bower M (2006) Opposing rules of dendritic cells subsets in HIV infection.  Blood 108: 1785-1786.
\bibitem{wk1}Wu L, Kewalramani V N (2006)  Dendritic-cell interactions with HIV: Infections and viral dissemination. Nat Rev Immunol 6:859-868.
\bibitem{mdh}Mara撲n C, Desoutter J F, Hoeffel G, Cohen W, Hanau D, Hosmalin A (2004) Dendritic cells cross-present HIV antigens from live as well as apoptotic infected CD4+ T lymphocytes. Proc. Natl. Acad. Sci. USA 101: 6092-6097.
\bibitem{h3p}Hlavacek W S, PercusJ K, Percus O E, Perelson A S, Wofsy C (2002) Retention of antigen on follicular dendritic cells and B lymphocyte through complement-mediated multivalent ligand-receptor interactions: Theory and implications to HIV treatment.  Math Biosci 176:185-202.
\bibitem{cune}McCune J M (2001) The dynamics of TCD4+ T cell depletion on HIV disease. Nature 410:974-979.
\bibitem{dpk}Douek D C, Picker L J, Koup R A (2003) Tcell dynamics in HIV-1 infection. Annu Rev Immunol 21:265-304.
\bibitem{mptr}Mohri H, Perelson A S, Tung K, Ribeiro R M, Ramratnam B et al. (2001) Increased turnover of T lymphocyte in HIV-1 infections and its reduction by antiretroviral therapy. J Exp Med 9:1277-1287.
\bibitem{psld}Porcheray F, Samah B, Leone C, Dereuddre-Bosquet N, Gras G (2006) Macrophage activation and human immunodeficiency virus infection: HIV replication directs macrophages toward a pro-inflammatory phenotype while previous activation modulates macrophage susceptibility to infection and viral production.  Virology 349: 112-120.
\bibitem{vs}Verotta D, Schaedeli F (2002) Non-linear dynamics models characterizing long-term virological data from AIDS clinical trials.  Math Biosci 176:163-83.
\bibitem{mcb}Massad E, Coutinho F A B, Burattini M N, Sallum P C, Lopez L F (2001)  A mixed ectoparasite-microparasite model for bat-transmitted rabies.  Theor Popul Biol 60: 265-279.
\bibitem{kp}Kim H, Perelson A S (2006) Viral and latent reservoir persistence in HIV-1 infected patients on therapy. PLoS Comput Biol 2(10):e135.doi:10.1371/journal.pcbi.0020135.
\bibitem{ael}Asquith B, Edwards C T T, Lipsitch M, McLean A R (2006) Inefficient cytotoxic T lymphocyte-mediated killing of HIV-1-infected cells in vivo. PLoS Biol 4:1-10.
\bibitem{smw}Schrier R D, McCutchen J A, Wiley C A (1993) Mechanism of immune activation of human immunodeficiency virus in monocytes/macrophages. J Virol 67:5713-5720.
\bibitem{slcg}Schacker T, Little S, Connick E, Gebhard-Mitcehll K, Zhang Z Q et al. (2000)  Rapid accumulation of human immunodeficiency virus (HIV) in lymphatic tissue reservoirs during acute and early HIV infections: Implications for timing of antiretroviral therapy. J Infect Dis 181:354-357.
\bibitem{mpbg}	Moris A, Pajot A, Blanchet F, Guivel-Benhassine F, Salcedo M et al. (2006) Dendritic cells and HIV-specific CD4+ Tcells: HIV antigen presentation, Tcell activation, and viral transfer. Blood 108:1643-1651.
\bibitem{haase}	Haase A T. Population biology of HIV infection: Viral and CD4+ T cell demographic and dynamics in lymphatic tissues.  Ann Rev Immunol 17:625-656.
\bibitem{rma}	Rouzine I M, Murali-Krishna K, Ahmed R (2005)  Generals die in friendly fire or modeling immune response to HIV.  J Comput Appl Math 184:258-274.
\bibitem{rossi}Rossi M, Lopez, L F (2009) The spread of HIV infection on immune system: Implications on cell populations and $R_0$ epidemic estimate.  Biomat 2009: International Symposium on Mathematical and Computational Biology. 1- 6 August 2009, Brasilia, Brazil; World Scientific, 331-341.
\bibitem{bh}De Boer R, Hogeweg P (1986) Interactions between macrophages and T-lymphocytes: Tumour sneaking through intrinsic to helper Tcell dynamics. J. Theor. Biol 120:331-351.
\bibitem{aelm}Asquith B, Edwards C T T, Lipsitch M, McLean A R (2006)  Inefficient cytotoxic T lymphocyte-mediated killing of HIV-1-infected cells in vivo.  PLoS Biol. 4:1-10.
\bibitem{dt}Duffin R P, Tullis R H (2002). Mathematical model that predict the complete course of HIV-1 infection and AIDS. J Theor Med. 4:157-166.
\bibitem{wk2}	Wigginton J E, Kirschner D E (2001) A model to predict cell-mediated immune regulatory mechanism during human infection with Mycobacterium tuberculosis. J Immunol.166:1951-1967.
\bibitem{kll}Kim P, Lee P, Levy D (2007) Modeling regulation mechanism in the immune system.  J Theor Biol. 246:36-69.

\end{thebibliography}



\end{document}